\documentclass{aastex}
\usepackage{spr-astr-addons}
\usepackage{url}

\RequirePackage{color}

\begin{document}

\title{Noether Gauge Symmetry Approach in $f(R)$ Gravity}

\shorttitle{Noether Gauge Symmetry Approach in $f(R)$ Gravity}
\shortauthors{Hussain et al.}

\author{Ibrar Hussain\altaffilmark{1}}
\and
\author{Mubasher Jamil\altaffilmark{2}}
\and
\author{F. M. Mahomed\altaffilmark{3}}

\altaffiltext{1}{School of Electrical Engineering and Computer
Science, National University of Sciences and Technology, H-12,
Islamabad, Pakistan. Email: ibrar.hussain@seecs.nust.edu.pk}

\altaffiltext{2}{Center for Advanced Mathematics and Physics,
National University of Sciences and Technology, Islamabad, Pakistan.
Email: mjamil@camp.nust.edu.pk}

\altaffiltext{3}{Centre for Differential Equations, Continuum
Mechanics and Applications,School of Computational and Applied
Mathematics, University of the Witwatersrand, Wits 2050, South
Africa. Email: Fazal.Mahomed@wits.ac.za}

\begin{abstract}

We discuss the $f(R)$ gravity model in which the origin of dark
energy is identified as a modification of gravity. The Noether
symmetry with gauge term is investigated for the $f(R)$ cosmological
model. By utilization of the Noether Gauge Symmetry (NGS) approach,
we obtain two exact forms $f(R)$ for which such symmetries exist.
Further it is shown that these forms of $f(R)$ are stable.

\end{abstract}


\section{Introduction}

From the observational data of supernovae of Type Ia (SN Ia)
\citep{riess,perl}, it was reported that the present observable
Universe is undergoing an accelerating phase. The mysterious source
for this late-time acceleration was dubbed `dark energy'. Despite
many years of research (see e.g., the reviews \citep{reviews,luca})
its origin has not been identified yet. Dark energy possesses
negative pressure leading the accelerated expansion of the Universe
by counteracting the gravitational force. One possible approach for
the construction of dark energy models is the modification of the
geometrical part of the Einstein equations. This approach known as
Modified Gravity can successfully explain the rotation curves of
galaxies, the motion of galaxy clusters, the Bullet Cluster, and
cosmological observations without the use of dark matter or
Einstein's cosmological constant \citep{mofat}.

The $f(R)$ theory of gravity is a candidate of modified theories of
gravity which is obtained by replacing the Ricci scalar $R$ with an
arbitrary function $f(R)$ in the Einstein-Hilbert Lagrangian (see
\citep{nojiri1,nojiri2} for reviews on this topic). Recently
different forms of $f(R)$ have been proposed \citep{noji,eli,cog}.
These theories can produce inflation, mimic behavior of dark matter
and current cosmic acceleration if suitable positive and negative
powers of curvature are added into the Einstein-Hilbert action
\citep{NO,noji,carr,Star}. Besides, compatibility with observational
data, the modified gravity theory has to be compatible with general
relativity and other viability conditions
\citep{noji1,noji2,luca,setare,shinji,jamil}.

The plan of this paper is as follows: In Section II, we discuss
briefly the basics of the $f(R)$ model and construct the modified
field equations. Then in Section III, we discuss the Noether gauge
symmetry approach for the $f(R)$ modified gravity. In Section IV, we
check the stability conditions on the $f(R)$ functions and finally
conclude our paper in the last Section. Throughout this paper, we
choose units $c=1=16\pi G$ and the metric signature $(-,+,+,+)$.

\section{Field equations in $f(R)$ theory of gravity}

In this section a spatially flat Friedman-Robertson-Walker (FRW)
Universe within the framework of $f(R)$ gravity is considered. Our
aim is to investigate models which exhibit Noether symmetry with
gauge term; we do not take in account matter contribution in the
action. We start with a (3+1)-dimensional action \citep{faro}
\begin{equation}\label{1}
S=\int d^4x\sqrt{-g}f(R),
\end{equation}
where $R$ is the scalar curvature and $f(R)$ is an arbitrary
non-linear function of $R$. The first model proposed had the form
$f(R) = R - \mu^4 /R$, in which the correction in $R^{-1}$ becomes
important only at low curvatures $R\rightarrow0$. It is demonstrated
in \citep{15,16} that action (1) is never renormalizable. In order
it to be renormalizable, as it is shown there, $L$ should consists
of $R^2+R_{\mu\nu}R^{\mu\nu}+R+\Lambda$. In addition to the desired
phenomenological properties of modified gravity in cosmology, there
is some motivation for these models from M-theory \citep{18}.

Variation of the action (1) with respect to the metric yields the
field equations
\begin{eqnarray}\label{2}
&&\frac{1}{2}g_{\mu\nu}f(R)-R_{\mu\nu}f'(R)+\nabla_\mu \nabla_\nu
f'(R)\nonumber\\&&-g_{\mu\nu}\Box f'(R)=0,
\end{eqnarray}
where a prime denotes derivative with respect to $R$ and $\mu$,
$\nu=0,1,2,3$. The operator $\nabla_\mu$ represents covariant
derivative and $\Box=\nabla^\mu\nabla_\mu$. We assume that the
geometry of the spacetime is given by the flat FRW line element
\begin{equation}\label{3}
ds^2=-dt^2+a^2(t)(dx^2+dy^2+dz^2).
\end{equation}
With the consideration of this model the field equations become
\begin{eqnarray}\label{4}
2\dot H + 3H^2&=&-\frac{1}{f'}\Big[ f'''\dot R ^2+f''(2H\dot R
+\ddot R)\nonumber\\&&+\frac{1}{2}(f-Rf') \Big],\\
H^2&=&\frac{1}{6f'}[f'R-f-6\dot RHf''],\label{5}
\end{eqnarray}
where $H=\dot{a}/a$ is the Hubble parameter and overdot denotes the
derivative with respect to proper time $t$. To investigate NGS an
effective Lagrangian for the model is needed whose variation with
respect to its dynamical variables yields the correct form of the
equations of motion. We follow the work of Souza and Faraoni
\citep{SF} and consider the above action that represents a dynamical
system in which the scale factor $a$ and the scalar curvature $R$
play the role of dynamical variables. The action (1) can be written
as \citep{capo2}
\begin{eqnarray}\label{6}
S&=&\int dt \mathcal{L}(a,\dot a,R,\dot R)=\int dt \Big[
a^3f(R)\nonumber\\&&-\lambda\Big\{ R-6\Big(H^2+\frac{\ddot
a}{a}\Big)  \Big\} \Big],
\end{eqnarray}
where the definition of $R$ is introduced in terms of $a$ and its
derivatives as a constraint. In order to apply the NGS approach, one
may easily verify that, in the FRW model, the Lagrangian related to
the above action (6) takes the form
\begin{equation}\label{7}
\mathcal{L}(a,\dot a,R,\dot R)=6(\dot a^2af'+\dot a\dot
Ra^2f'')+a^3(f'R-f).
\end{equation}
Varying the Lagrangian (7) with respect to $R$ yields the following
relation for the scalar curvature
\begin{equation}\label{8}
R=6\Big(H^2+\frac{\ddot a}{a}\Big).
\end{equation}
The equation (4) can be obtained by varying the Lagrangian (7) with
respect to $a$.

\section{Noether gauge symmetries in $f(R)$ theory of gravity}

Noether symmetries are the symmetries of the Lagrangians which have
found a recent impetus since these can reveal new features of the
gravitational theories \citep{capo2,capo3,capo4,capo5}. The NS are
essential tools for solving the gravitational field equations
\citep{sanyal,jamil}. In Scalar-Tensor cosmology, this approach
results in an extra correction term $R^{-1}$ and fixes the form of
the coupling parameter and the field potential \citep{capo}. This
approach yields an exact form of $f(R)$ functions relevant for
cosmological model \citep{babak,babak1,babak2,capo1,capo11,roshan}.
In the literature, Noether symmetries have been studied in the
context of $f(R)$ theory of gravity by ignoring the gauge function
in the Noether symmetry condition
\citep{babak,babak1,babak2,capo1,capo11}. We consider the gauge term
 of the Noether symmetries \citep{Ibr,blu}.  Here we apply the approach
of NGS to look at some interesting forms of $f(R)$.

A vector field
\begin{equation}
\mathbf{X}=\xi(t,a,R)\frac{\partial }{\partial
t}+\eta(t,a,R)\frac{\partial }{\partial
a}+\beta(t,a,R)\frac{\partial }{\partial R},
\end{equation}
whose first prolongation is
\begin{equation}\label{9}
\mathbf{X^{[1]}}=\mathbf{X}+\dot \eta(t,a,R)\frac{\partial
}{\partial \dot a}+\dot \beta(t,a,R)\frac{\partial }{\partial \dot
R},
\end{equation}
where
\begin{eqnarray}\label{10}
\dot\eta&\equiv&\frac{\partial \eta}{\partial t}+\dot
a(\frac{\partial\eta}{\partial a}-\frac{\partial \xi}{\partial
t})+\dot R\frac{\partial \eta}{\partial R}-\dot a^2\frac{\partial
\xi}{\partial a}-\dot a\dot R\frac{\partial \xi}{\partial R},\nonumber\\
\dot\beta&\equiv&\frac{\partial \beta}{\partial t}+\dot a
\frac{\partial \beta}{\partial a}+ \dot R(\frac{\partial\beta
}{\partial R}-\frac{\partial \xi}{\partial t})-\dot
R^2\frac{\partial \xi}{\partial R}\nonumber\\&&-\dot a\dot
R\frac{\partial \xi}{\partial a},
\end{eqnarray}
is called a NGS if the following condition holds
\begin{equation}\label{11}
\mathbf{X^{[1]}}\mathcal{L}+(\mathbf{D}\xi)\mathcal{L}=\mathbf{D}A(t,a,R).
\end{equation}
Here $A$ is the gauge function and
\begin{equation}\label{12}
\mathbf{D}\equiv\frac{\partial }{\partial t}+\dot a\frac{\partial
}{\partial a}+ \dot R\frac{\partial}{\partial R}.
\end{equation}

Using the Lagrangian (7) in (12) and after the separation of
monomials we obtain the following system of determining equations
\begin{equation}
\xi_{,a}=0,\ \ \ \xi_{,R}=0,\label{13}
\end{equation}
\begin{equation}
\eta f'+\beta a
f''+2af'\eta_{,a}-\xi_{,t}af'+a^2f''\beta_{,a}=0,\label{14}
\end{equation}
\begin{equation}
12af'\eta_{,t}+6a^2f''\beta_{,t}=A_{,a},\label{15}
\end{equation}
\begin{equation}
f''\eta_{,R}=0,\label{16}
\end{equation}
\begin{equation}
6a^2f''\eta_{,t}=A_{,R},\label{17}
\end{equation}
\begin{equation}
2af''\eta+a^2f'''\beta+a^2f''(\eta_{,a}+\beta_{,R}-\xi_{,t}) +
2af'\eta_{,R}=0,\label{18}
\end{equation}
\begin{equation}
a^2(3\eta+a\xi_{,t})(f'R-f)+a^3f''R\beta=A_{,t},\label{19}
\end{equation}
where ``," denotes partial derivative.

From (\ref{16}), for $f''=0$, we get $\eta_{,R}\neq0$, then from
(\ref{18}), $f'=0$. Therefore the two functions $\xi$ and $A$ become
only functions of the variable $t$ and (\ref{19}) reduces to
\begin{equation}
a^2(3\eta+a\xi_{,t})f=A_{,t}.
\end{equation}
Differentiating this last equation with respect to $R$ we get
\begin{equation}
\eta_{,R}=0,
\end{equation}
which leads to a contradiction. Hence we take $f''\neq0$, so from
(\ref{16})
\begin{equation}\label{21}
\eta_{,R}=0.
\end{equation}

There are three cases in the solution of the above determining
equations (14)-(20). These are given below.

{\bf Case-I}. If $f$ is arbitrary, the solution of the above system
(14)-(20) gives rise to the following NGS generator
\begin{equation}
{\bf X}=\frac{\partial}{\partial t},\label{1}
\end{equation}
and the gauge term is a constant which can be taken as zero. The
energy type first integral is
\begin{equation}
I=6\dot a^2 a f'+6\dot a\dot R a^2 f''- a^3(f'R-f).\label{1a}
\end{equation}

{\bf Case-II}. If $f$ is the fractional power law, viz.
\begin{equation}
f=f_0 R^{3/2},\label{2n}
\end{equation}
then the solution of the determining equations (14)-(20) yields
\begin{eqnarray}
\xi=b_1t+b_2,\nonumber\\
\eta=\frac23 b_1a+ b_3 a^{-1},\nonumber\\
\beta=-2R(b_1+ b_3a^{-2}).\label{3}
\end{eqnarray}
The gauge term is zero. The corresponding generators are
\begin{eqnarray}\label{22}
\textbf{X}_0&=&\frac{\partial}{\partial t},\nonumber\\
\textbf{X}_1&=&t\frac{\partial}{\partial t}+\frac23
a\frac{\partial}{\partial a} -2R\frac{\partial}{\partial
R},\nonumber\\
\textbf{X}_2&=&a^{-1}\frac{\partial}{\partial a}-2R
a^{-2}\frac{\partial}{\partial R},
\end{eqnarray}
which constitute a well-known three-dimensional algebra with
commutation relations
\begin{equation}\nonumber
 [\textbf{X}_0,\textbf{X}_1]=0,\,
[\textbf{X}_0, \textbf{X}_2]=0,\, [\textbf{X}_1,
\textbf{X}_2]=-\frac43 \textbf{X}_2.
\end{equation}
Here too the gauge term is zero. The corresponding first integrals
are
\begin{eqnarray}\label{2a}
I_0&=&9\dot a^2 a f_0R^{1/2}+\frac92\dot a\dot R a^2 f_0R^{-1/2}- \frac12
a^3f_0R^{3/2},\nonumber\\
I_1&=&-9\dot a^2 a tf_0R^{1/2}+\frac12f_0ta^3R^{3/2}+3\dot a
a^2f_0R^{1/2}\nonumber\\&&
+3\dot Ra^3f_0 R^{-1/2}-\frac92\dot a\dot R ta^2f_0R^{-1/2},\nonumber\\
I_2&=&9\dot a f_0R^{1/2}+\frac92 a f_0R^{-1/2}\dot R.
\end{eqnarray}

{\bf Case-III}. If $f$ is a general power law of the form
\citep{capo,capo1,capo2,capo3,capo4,capo5}
\begin{equation}
f=f_0R^{\nu},\quad \nu\ne 0,1,\frac{3}{2},\label{5n}
\end{equation}
then the above linear determining equation system has solution
\begin{equation}
\xi=b_1t+b_2,\ \ \eta=\frac{2\nu-1}{3}a b_1,\ \
\beta=-2Rb_1.\label{6}
\end{equation}
Here the gauge term is zero and $f_0$ is a constant in (30).

Note that if $\nu=0,1,\frac{3}{2}$, then $f$ is a constant, linear
or fractional power law of Case-II. These are thus excluded in
Case-III. It should also be stated that for Case-III there are two
Noether symmetries. The Noether symmetry generators are given by
\begin{equation}\label{23}
\textbf{X}_0=\frac{\partial}{\partial t},\ \
\textbf{X}_1=t\frac{\partial}{\partial
t}+\frac{2\nu-1}{3}a\frac{\partial}{\partial a}
-2R\frac{\partial}{\partial R}.
\end{equation}
The Lie algebra is the two-dimensional Abelian Lie algebra,
$[\textbf{X}_0,\textbf{X}_1]=0$. We note that for the fractional
power law the algebra is three-dimensional and for the arbitrary
power law it is two-dimensional and thus the symmetry breaks by one.

The corresponding first integrals for this case are
\begin{eqnarray}\label{3a}
I_0&=&6\dot a^2 a\nu f_0R^{\nu-1}+6\dot a\dot R
a^2\nu(\nu-1)f_0R^{\nu-2}
\nonumber\\&&-a^3f_0R^{\nu}(\nu-1),\nonumber\\
I_1&=&-6\dot a^2 atf_0\nu
R^{\nu-1}+a^3tf_0(\nu-1)R^{\nu}\nonumber\\&& +4\nu(2-\nu)\dot
aa^2f_0R^{\nu-1}+2\nu(\nu-1)(2\nu-1)\nonumber\\
&&\times\dot Ra^3f_0R^{\nu-2}-6\nu(\nu-1)\dot a\dot
Rta^2f_0R^{\nu-2}.
\end{eqnarray}

Hence we have obtained some different $f(R)$ functions from our
analysis. In the next section, we check the conditions under which
our $f(R)$ functions can produce viable cosmology.

\section{Stability Analysis}

The stability conditions for $f(R)$ gravity are
\citep{luca,noji,staro}
\begin{enumerate}
  \item $f'>0$ for $R\geq R_0$, where $R_0>0$ is the Ricci
  scalar at the present epoch. This is also required to avoid
  anti-gravity.
  \item $f''>0$ for $R\geq R_0$. It ensures consistency with local
  gravity tests, presence of the matter dominated epoch and the
  stability of cosmological perturbations.
  \item $f(R)\rightarrow R-2\Lambda$ for $R\gg R_0$. This is
  required for the local gravity tests and for the presence of the
  matter dominated epoch.
  \item $0<\frac{Rf''}{f'}(r=-2)<1$ at $r=\frac{-Rf'}{f}=-2$. This
  is required for the late-time de Sitter point.
\end{enumerate}

To ensure classical and quantum stability in the region $R$, we want
our $f(R)$ theory to satisfy conditions (1) and (2). The first
condition means that gravity is attractive and the graviton is not a
ghost. It was recognized long ago that its violation during the time
evolution of a FRW background results in the immediate loss of
homogeneity and isotropy and formation of a strong space-like
anisotropic curvature singularity \citep{10}. The second condition
is necessary so that in the case of $f(R)$ models of present dark
energy, the necessity to keep it valid for all values of $R$ during
the matter- and radiation-dominated epochs in order to avoid the
Dolgov-Kawasaki instability \citep{noii,nojj,faro1}. Condition (3)
imposes that the modified $f(R)$ gravity must reduce to Einstein
gravity under a suitable limit of the curvature scalar. Finally
condition (4) tells us of a de Sitter point which corresponds to a
vacuum solution with constant $R$ i.e. $\Box f'=0$ at this point. It
trivially implies that any quadratic $f(R)\sim R^2$ will satisfy
this condition and gives rise an exact de Sitter solution. Hence
quadratic $f(R)$ functions are particularly useful to model
inflation.

The stability conditions are satisfied in the following manner
respectively:
\begin{itemize}
\item Conditions (1) and (3) give $f'=f_0\frac{3}{2}\sqrt{R}>0$ and
$f''=f_0 \frac{3}{4\sqrt{R}}>0$ since $R$ is positive. Condition (4)
is not valid in this case. Condition (5) yields $0<\frac{1}{2}<1$ at
$r=-2$.

\item First two conditions hold $f'=f_0\nu R^{\nu-1}>0$, $f''=f_0\nu (\nu-1)R^{\nu-2}>0$
whenever $f_0>0$, $\nu>1$. Condition (3) is not valid here since
$\Lambda=0$ and $\nu\neq1$. Condition (4) is invalid: it implies
$0<(\nu-1)<1$ at $r=-2$ but $\nu>1$.
\end{itemize}

\section{Conclusion}

In this paper we have considered the 4-dimensional $f(R)$ theory of
gravity. We have taken the spatially flat FRW Universe in the
framework of $f(R)$ gravity. Our aim was to investigate models which
exhibit Noether symmetry with {\it gauge} term, where we have
ignored the matter contribution in the action. On using the
Lagrangian (7) in the definition of NGS (10) we have obtained a
system of partial differential equations (14) - (20), which involve
four unknown functions $\xi$, $\eta$, $\beta$ and $A$, where each is
a function of three variables $t$, $a$ and $R$. The integration of
the equations (14) - (20) has yielded two different solutions
(\ref{2n}) and (\ref{5n}). Solving the system of equations, we
obtained the gauge function to be zero. We then constructed the
symmetry generators and the corresponding conserved quantities (also
known as first integrals or constants of motion). We note that for
the fractional power law, the algebra is three-dimensional and for
the arbitrary power law, it is two-dimensional and thus the symmetry
breaks by one. A previous study of ours \citep{jamil} discussing the
$f(R)$-tachyon model also resulted a zero gauge function. Thus we
conjecture that the application of NGS to generic $f(R)$ Lagrangian
results in zero gauge function. The symmetry generators give the
time translational and scaling symmetry of the theory. The stability
analysis of the two forms of $f(R)$ obtained here has been done in
Section IV. We have shown that the two forms of $f(R)$ given by
(\ref{2n}) and (\ref{5n}) are stable and consistent with the local
gravity tests by imposing some restrictions on the constants
involved. Recently some black hole solutions have been studied in
$f(R)$ theory of gravity using some known forms of $f(R)$
\citep{myung1,myung2,myung3,habib}. It would be of interest to study
black hole solutions using the forms of $f(R)$ obtained here.

\section*{Acknowledgment} The authors would deeply thank the
referee for giving very insightful comments on our work.

\end{document}